\documentclass[twocolumn,
prd,showpacs,floats,amsmath,amssymb]{revtex4}
\usepackage{epsfig}
\usepackage{amsmath}
\renewcommand{\(}{\left(}
\renewcommand{\)}{\right )}
\renewcommand{\[}{\left [}
\renewcommand{\]}{\right ]}
\def\fslash#1{#1 \!\!\! \slash}
\def\beq{\begin{equation}}
\def\eeq{\end{equation}}
\def\pa{\partial}

\def\bea{\arraycolsep .1em \begin{eqnarray}}
\def\eea{\end{eqnarray}}

\def\Tr{{\rm Tr}}

\let\si=\sigma

\let\no=\nonumber

\def\et{{\em et al.}}
\def\eq#1{Eq.(\ref{#1})}
\def\refr#1{\cite{#1}}

\def\eqs#1{Eqs.(\ref{#1})}
\def\s0#1#2{\mbox{\small{$ \frac{#1}{#2} $}}}
\def\0#1#2{\frac{#1}{#2}}

\def\anp#1#2#3{Adv.\ Nucl.\ Phys. \ {\bf #1}, #2  (#3)}
\def\plb#1#2#3{Phys. Lett. {\bf B #1}, #2  (#3)}
\def\npa#1#2#3{Nucl. Phys. {\bf A #1}, #2  (#3)}
\def\npb#1#2#3{Nucl. Phys. {\bf B #1}, #2  (#3)}
\def\ppnp#1#2#3{Prog. Part. Nucl. Phys. {\bf #1}, #2  (#3)}

\def\pra#1#2#3{Phys. Rev.  {\bf A #1}, #2  (#3)}

\def\prc#1#2#3{Phys. Rev.  {\bf C #1}, #2  (#3)}
\def\prd#1#2#3{Phys. Rev. {\bf D #1}, #2  (#3)}

\def\prl#1#2#3{Phys. Rev. Lett. {\bf #1}, #2  (#3)}
\def\phys#1#2#3{Physica {\bf #1}, #2  (#3)}
\def\ann#1#2#3{Ann. Phys. (N.Y.) {\bf #1}, #2  (#3)}
\def\anp#1#2#3{Adv. Nucl. Phys. {\bf #1}, #2  (#3)}

\def\ijmpa#1#2#3{Int.\ J.\ Mod.\ Phys.\ {\bf A #1}, #2  (#3)}
\def\mpla#1#2#3{\ Mod.\ Phys.\ Lett. {\bf A #1}, #2  (#3)}
\def\ijmpe#1#2#3{Int.\ J.\ Mod.\ Phys.\ {\bf E #1}, #2  (#3)}
\def\jhep#1#2#3{J.\ High Energy Phys.\ {\bf #1}, #2  (#3)}
\def\rmp#1#2#3{Rev.\ Mod.\ Phys.\ {\bf #1}, #2  (#3)}
\def\zpa#1#2#3{Z.\ Phys.\ {\bf A #1}, #2  (#3)}

\def\ibid#1#2#3{{\it ibid.}, {\bf #1}, #2  (#3)}

\begin{document}
\title{
Unusual photon isospin mixing and instantaneous Coulomb
    effects on the thermodynamics of compact matter}
\author{Ji-sheng Chen\footnote{chenjs@iopp.ccnu.edu.cn}}
\affiliation{
Institute of Particle Physics $\&$ Physics Department, Hua-Zhong
Normal University, Wuhan 430079, People's Republic of China}
\begin{abstract}
A hidden local symmetry formalism with a two-photon counterterm
    approach is performed based on the relativistic
    continuum quantum many-body theory.
The underlying electromagnetic under-screening as well as
    screening effects between the electric charged point-like
    electrons and composite protons are discussed by analyzing the
    in-medium isospin mixing of Lorentz vector with scalar due to electromagnetic photon.
Besides the usual screening results, the main conclusion is that
    an effective oscillatory instantaneous Coulomb potential between
    the like-charged collective electrons
    contributes a very large negative term to the equation of
    state.
This counterintuitive like-charged attraction results
    from the modulation factor of the opposite charged baryon superfluid background.
The anomalous long distance quantum dragging effects between the
    collective electrons can be induced in a compact Coulomb confinement
    environment. The physics is of the strongly coupling characteristic in a specific dilute regime.\end{abstract}
 \maketitle
\section{Introduction}
Intriguingly, it is usually assumed that there is a uniform
    positive background to ensure the stability in discussing the
    thermodynamics of electrons system.
This is the well-known (generalized) Thomson Problem for over a
    century.
Of course, there is not a finished standard answer up to
    now\refr{Thomson}.
In nuclear physics, the status quo is an
    \textit{inverse} one.
The electrons are usually taken as a uniform
    negative charged background and further treated as the ideal
    degenerate gas with the notion that the nuclear force is of
    electric charge independent.
Furthermore, it is customarily accepted that the weak Coulomb
    contribution is well known or screened out and can be turned off
    by taking the point-like picture of protons.
What physics will be found or answered if considering the
    interactions between the two opposite charged subsystems?

The realistic nuclear matter is subject to the long range
    electromagnetic(EM) interaction.
The infrared EM Coulomb force between the charged particles such
    as electrons and composite protons makes the discussion of
    thermodynamics or equation of states(EOS) even more involved and
    may lead to rich physics\refr{gulminelli2005,Magierski,Maruyama2004,Horowitz2005}.
In addition to the (residual) strong interaction contribution to
    EOS, the EM fluctuation/correlation effects are very important to
    the mechanical, thermal, chemical equilibrium conditions and
    further influence the various transport properties of the compact
    star nuclear matter due to Coulomb ``frustration" effects (for
    example, see Refs. \refr{Horowitz2005} and references
    therein).
This kind of frustration effects will lead to the strongly coupled
picture physics which belongs to the  non-perturbative context.
The hitherto overlooked but important EM force can even
    significantly affect the single particle energy spectrum as
    recently found with the study of some mirror nuclei,
which provides evidence for the inclusion of the novel
    electromagnetic spin-orbit effect\refr{Ekman2004}.
To our knowledge, the EM correlation is still a challenging task
    in many-body theory and even for finite nuclei\refr{bulgac1999}.

The effective relativistic continuum nuclear theory provides a
    powerful theoretical framework to account for the various exotic
    many-body properties of neutron-star nuclear matter or finite
    nuclei\refr{walecka1974,Ring1996}.
The pairing correlation calculations of neutrons and protons in
    nuclear matter inspired by the quite different empirical negative
    scattering lengths in free space between nucleons promote us to
    assume the spontaneous breaking analogously to the EM $U(1)$ gauge
    symmetry breaking concept of Anderson-Higgs mechanism in order to
    gauge the infrared singularity\refr{ring1991,chen2005}.
This is the Lorentz spontaneous violation according to the
    standard notion, which is currently the central topic of cosmology
    etc.\refr{jenkins2004,Carroll}.
The theoretical spinodal instability and uncertainty can be
    removed consequently with the in-medium Lorentz violation notion.

The electron density becomes very diluter(hence strongly
    correlating according to statistical physics) compared with the
    baryon density when one takes into account the charge neutralized
    condition\refr{chen2005}.
It remains an intriguing task to deal with the strongly correlated
    electrons ground state energy simultaneously with that of
    nucleons, which is closely related with the neutron star crust or
    supernova explosion physics in the dilute regime.
Needless to say, this is a fundamental/difficult problem to tackle
    analytically in the general many-body context.

To address the charge neutralized strongly interacting
    system thermodynamics with the net electric charged chemical potential based on the relativistic continuum field
    theory formalism,
    we propose a compensatory two-photon method with the Hidden Local
    Symmetry(HLS) formalism to remove the infrared singularity.
Surprisingly, the exact ground state energy solution of collective
    electrons manifests a counterintuive long range attraction picture
    in the strongly coupling limit,
    which is far from the starting
    point in order to unveil/answer the realistic physics motivation
    presented in Ref.\refr{chen2005}.

The plan of this work is as follows. In Section \ref{sec2},
    the isospin mixing effect of photon with sigma meson(short range nuclear
    force) leading to under-screening is analyzed.
The consequent novel quantum fluctuation effects on the
    strongly correlated electrons thermodynamics
    in a dilute compact confinement environment is discussed in Section \ref{sec3}.
The discussions are made in Section \ref{sec4}. The final
Section\ref{sec5} presents the conclusion remarks.
\section{Isospin mixing of photon with sigma meson}\label{sec2}

To make the constructive two-photon Thomson Problem counterterm
    scheme presented in the next section more comprehensive,
    we will firstly reveal the unusual polarization properties that are often overlooked.
The discussions will give us a strong
    inspiration, i.e.,
    there is the possible EM under(anti)-screening effects
    in addition to the usual screening in a compact environment.
The expected conventional weak coupling perturbative expansion
and/or resummation based on the exclusive screening of
    normal state cannot be valid any more in addressing the strongly coupling limit thermodynamics.

In addressing the low energy long wavelength thermodynamics,
it is
    a good approximation to take the nucleon as a point-like particle
    according to the effective field theory's idea\refr{walecka1974}.
With this picture, one can perform the polarization tensor
    calculation with the in-medium protons\refr{Leinson2000}.
Essentially, according to particle physics, one can introduce the
    isospin concept for photon $\gamma $.
Therefore we can study
    the nontrivial in-medium isospin mixing of $\gamma$ and sigma meson $\sigma$ following
    the well discussed $\sigma-\omega$ mixing etc.,
    from which the full propagators of $\gamma $  and $\sigma $ can be simultaneously
    determined with the mixing polarization tensor $\Pi$,
    a $5\times 5 $ matrix\refr{Chin1977}.

With $\Pi $, the
    dielectric function in the presence of mixing to determine the
    collective modes or study the screening effects is
\bea
 \epsilon (k_0,|{\bf k}|) =\epsilon ^2_T \times \epsilon_{mix},
\eea where $\epsilon _T $ corresponds to the two identical
transverse (T) modes
    and $\epsilon_{mix}$ to the longitudinal mode (L) of the photon with mixing.
Furthermore, one can analyze the nontrivial mixing effects on the
inter-particle EM correlation physics between the charged
particles with the effective permittivity, which is reduced to
\bea\label{mixing}
    &&\epsilon _{eff}(0,|{\bf k}|\rightarrow 0 )\no\\
    &&~~~=1+\01{{\bf k}^2}\(\Pi^L_P(0,0)+\01{m_\sigma^2+\Pi^\sigma_{N,P}}(\Pi_P^{\sigma-\gamma,0})^2\).
\eea In \eq{mixing}, the polarization tensor components are
calculated via the following selfenergies
 \bea\label{self-energy}
    \Pi ^\si _{N,P}(k)&&=2 g_\si ^2 T \sum _{p_0}\int_p
        \Tr
        \[\0{1}{\fslash{p}-M^*}\0{1}{(\fslash{p}-\fslash{k})-M^*}\],\no\\
   \Pi ^{\sigma-\gamma, \mu}_{P}(k)&&=eg_\si T \sum _{p_0} \int_p \Tr
        \[ \0{1}{\fslash{p}-M^*}\gamma
        ^\mu\0{1}{(\fslash{p}-\fslash{k})-M^*}\], \\
   \Pi ^{\gamma, \mu\nu}_{P}(k)&&=e^2T\sum _{p_0}\int_p \Tr
        \[\gamma ^\mu \0{1}{\fslash{p}-M^*}\gamma
        ^\nu\0{1}{(\fslash{p}-\fslash{k})-M^*}\],\no
\eea
with the shorthand notation $\int _p=\int {d^3{\bf p}}/{(2\pi
    )^3}$ and $p_0=(2n+1) \pi T i+\mu ^*$ while $\mu^*$ and $M^*$
    being the effective chemical potential and fermion mass, respectively.
The $\Pi ^{\gamma, \mu\nu}_{e} $ for the electron-loop self-energy
    is similar to $\Pi ^{\gamma,\mu\nu}_{P}$.

The standard calculations of the self-energies including the Dirac
and Fermi's sea contribution at finite temperature can be found in
Ref.\refr{chen2002}. Here in this work, concentrating on the
density fluctuation effects and relevant
    physics, we present explicitly the Fermi sea's contribution of scalar meson self-energy and the $\sigma -\gamma $ mixing polarization tensor-$0$ component at $T=0$
\bea
    \Pi ^{\sigma-\gamma,0}_{P}(0, 0)
    &&=-\0{e g_\sigma M^*}{\pi^2} k_f,\no\\\label{important1}
    \Pi ^{\sigma}_{N,P}(0, 0)
    &&=\0{g^2_\sigma }{2\pi^2} \[ k_f E_f^*\right. \left.+3 M^{*2}\ln\0{M^*}{k_f+E_f^*}\],
\eea where $E_f^*=\sqrt{k_f^2+M^{*2}}$. From \eq{important1}, one
can find
    that the static limit of the sigma meson self-energy is
    negative $\sim - 10^6 MeV^2$ in a wide density and temperature
    regime enhanced by the isospin degenerate factor ``2" in \eq{self-energy} due to the doping effects of neutrons.
This can be attributed to the
    different isospin vector and scalar fluctuation behaviors of
    QCD\refr{chen2002}.

The standard Thomas-Fermi/Debye screening masses are closely
    associated with EOS, which are defined according to the static
    infrared limit of the relevant self-energies\refr{kapusta1989}.
The static limit of the e(P)-loop self-energy is the familiar
    formalism
\bea\label{electron-mass1} \Pi ^{00}(0, 0)=-\0{e^2}{\pi
    ^2}k_f E_f^*.\eea
If without considering the non-trivial mixing effect,
    the usual electric screening masses reflecting the weak coupling screening
    physics can be directly obtained through
    $M_D^2=\Pi^L(0,0)=-\Pi^{00}(0,0)$.
For example, at the empirical saturation density with
    Fermi-momentum $k_f=256.1 MeV$ and an effective nucleon mass
    $M^*\sim 0.8 M_N$,
    the numerical results are $M_D^P\sim 43.46 MeV$ and
\bea\label{electron-mass}
    M_D^e=24.69 MeV.
\eea

The crucial observation is that the Thomas-Fermi electric
    screening mass  through the virtual
    (anti-)electron polarization calculation \eq{electron-mass1}-\eq{electron-mass} instead of that from the random phase
    approximation(RPA) P-loop one is exactly consistent
    with the previous estimation\refr{chen2005}.
This scenario is also consistent
    with the conventional notion of plasma theory that electrons screen the long-range Coulomb force
    between protons or ions\refr{Maruyama2004}.
The either local or global charge
    neutralized condition is automatically ensured through the full
    photon propagator calculation\refr{note2}.

In turn, a natural but very uncomfortable question may arise from
    above fortuitous but exact consistency
    Eqs.(\ref{electron-mass1}-\ref{electron-mass}) is:
Will the nucleons ``screen" the EM interaction between the
    opposite charged lepton electrons?

The systematic study is non-trivial as naively expected because
    the solutions will involve the Debye mass that can become
    catastrophically
    imaginary. With the denominator in the bracket of \eq{mixing},
    the term
    $\Pi^L_P(0,0)+(\Pi_P^{\sigma-\gamma,0})^2/({m_\sigma^2+\Pi_\sigma})
    $ may be smaller than $\Pi^L_P(0,0)$ and the effective
    permittivity can be negative due to the isospin mixing
    effects of isoscalar $\sigma$ and $\gamma$ with the unusual sigma
    meson self-energy behavior \eq{important1}.
In deed, the Fermi-sea contribution dominates over the Dirac
    sea's, i.e., the in-medium/density fluctuation effects can lead to
    the under(anti)-screening effects\refr{chen2002}.
The negative permittivity implies that there is a strongly coupled
    characteristic EM instability in a compact environment.
This Fermi surface instability results from the rummy competition
    between the short range nuclear force attraction and the long
    range EM fluctuation repulsion.

In our perception, the complete isospin mixing calculations for
    the strongly coupled nucleons is a seriously difficult task, which
    we suspect will require a greater understanding of nuclear force.
We do not claim to have a complete solution. Intuitively we can
make an {\em ansatz} that the final analytical
    expression can be reduced to
\bea
\Pi^L_P(0, 0) + \mbox{mixing contribution} =
        -\Pi^L_e(\mu_P^*, T_P),
\eea
which can be in principle realized
    through other involved isospin mixing discussions $\sigma-\omega$, $\gamma-\omega$ esp. the unclear charge
    symmetry breaking term $\rho-\omega$ etc. attributed to the
    different current quark masses which is beyond the scope of the adopted effective theory framework.
These effects
    can be manifested through the two-photon \refr{Evslin2005} approach in an easily controllable way.
The two-photon formalism as given in Sec. \ref{sec3} will
    confirm this ansatz indirectly through the Thomson stability condition
     of the lepton electrons thermodynamics background.

With excitation concepts in plasma dielectric function theory, the
    modes determined by the poles of the full propagators may be
    spacelike and of tachyonic characteristic which is closely related
    to the Landau damping and consistent with the nonlinear response
    property.
From the viewpoint of isospin mixing, the EM
    interactions between the point-like particles explored by the
    relativistic nuclear field theory are not simply/only screened and
    their contribution to thermodynamics cannot be neglected by
    naively taking the ideal degenerate gas or quasi-particle picture
    of normal fluid.
Resulting from long range Coulomb frustration effects, the
multi-component baryon nucleons system can be in the novel
thermodynamics pasta state especially in the subsaturation nuclear
density regime. The analytical exploration contributes to
understanding the underlying strongly interacting Coulomb physics
hidden in the diversely updating works.

\section{Strongly correlated compact matter
thermodynamics}\label{sec3}
\subsection{Formalism}

With above analysis, the isoscalar EM interaction embedded in the
relativistic nuclear
    theory has the strongly coupling confinement characteristic which
    promotes us to construct a Proca-like Lagrangian to deal with the
    unusual quantum mechanical Coulomb effects between the charged
    baryon particles\refr{chen2005}.
In this work, we attempt a two-photon non-perturbative
    approach to model
    the thermodynamics properties of the charge neutralized $Pe\nu N$
    plasma with global vanishing net electric charge chemical
    potential.
The effective theory involves the interaction of Dirac nucleons
    and electrons with an auxiliary Maxwell-Proca like EM field as
    well as the scalar/vector mesons fields
\bea
    {\cal L}_{QHD-\sigma,\omega,\rho}+{\cal L}_e;\no\\
    {\cal L}_{\gamma, free}= &&-\014 F_{\mu\nu} F^{\mu\nu};\no\\
    {\cal L}_{I, \gamma -P}=&&\012 m_{\gamma, e}^2 \delta _{\mu 0}A_\mu A^\mu
        + A_\mu J^\mu _P;\no\\\label{electron-current}
    {\cal L}_{I, \gamma -e}=&&\012 m_{\gamma,P}^2 \delta _{\mu 0}A_\mu A^\mu +A_\mu J^\mu _e.
\eea The full description for the relevant degrees of freedom in
    the Lagrangian can be found in
    the literature\refr{walecka1974,chen2005}.
In \eqs{electron-current}, the electric currents contributed by
    the baryon protons and lepton electrons are as follows,
    respectively
\bea
    && J^\mu _P= -e {\bar \psi _P} \gamma ^\mu
    \0{(1+\tau _3) }{2} \psi _P ,~~~~~ J^\mu _e= e {\bar l _e} \gamma ^\mu l_e.
\eea

Based on the local gauge invariant free Lagrangian, the gauge
    invariant effective interaction actions can be constructed by the
    standard way\refr{peskin1995,Dvali2005}, respectively
\bea
    {\cal L}_I=-\014 F_{\mu\nu}F^{\mu\nu}+|D_\mu H|^2+V(H)+A_\mu
    J^\mu.
\eea The two photon mass gaps(i.e., the Higgs charge and the
expectation value of the Higgs background field) are free Lagrange
multiplier parameters, i.e., they should be determined by relevant
thermodynamics identities and critical dynamics constraint
criterions.

Based on introducing the photon
    isospin concept,
    the conventional electromagnetic current is
    divided into two parts: one is the baryon current contribution of
    electric charged protons; the other is the lepton current of
    opposite charged electrons.
The position permutation of the two mass gap parameters in
    relevant effective actions \eq{electron-current} makes it possible
    to deal with the interior EM interaction as
    auxiliary external field approximation one another to ensure the
    theoretical thermodynamics self-consistency for the two opposite charged subsystems.

\subsection{Thermodynamics potential of strongly coupled electrons through vector condensation} The current conservation is
    guaranteed by the Lorentz transversality condition with HLS
    formalism
\bea
    \pa _\mu A^\mu=0,\eea
which can be naturally realized by taking the relativistic Hartree
    instantaneous approximation (RHA) formalism.

As a complementarity to the discussion for the nucleons background
thermodynamics\refr{chen2005} with the Coulomb frustrating
effects, we presently limit to the thermodynamics of the strongly
correlated dilute electrons (lepton) background in a
    dense and hot compact nuclear environment with the following effective Lagrangian
\bea\label{original}
    {\cal L}_{\textsl{effective}}=&&{\bar \psi_e} (i \gamma _\mu\pa ^\mu+\gamma _0 \mu _e -m_e) \psi_e\no\\
    && -\014 F_{\mu\nu} F^{\mu\nu}+\012 m_{\gamma, P}^2 A_\mu A^\mu+A_\mu
    J^\mu_e \no\\
    &&+\delta {\cal L}_{\textsl{Auxiliary~~Nuceons~Counterterm~Background}}.
\end{eqnarray} The $A_\mu$ is the vector field with the field stress
\bea
    F_{\mu\nu}=\pa_\mu A_\nu-\pa_\nu A_\mu.
\eea

In \eq{original}, $m_e$ is the bare electron mass with the
background fluctuating vector
    boson mass gap squared $m_{\gamma, P}^2$. In terms of finite
    temperature field theory with functional path
    integral\refr{walecka1974,kapusta1989,chen2005}, the effective
    potential reads
\bea
   \label{potential}\Omega_{e}/V
       =&&-\012 m_{\gamma,P}^2 A_0 ^2 -2T  \int _k
        \{\ln (1+e^{-\beta (E_e-\mu_e^*)}) \no\\
        &&~~~~~~~~~~+\ln (1+e^{-\beta (E_e+\mu_e^*)})\}.
\eea The particle(charge) number density $\rho _e=2\int _k
(n_e-{\bar n_e})$ is well defined by the \textit{thermodynamics
    identity}\bea\label{thermal}
\0{\pa \Omega _e}{\pa \mu
    _e}|_{A_0}\equiv-\rho _e.\eea
The tadpole diagram with photon
    self-energy for the full fermion electron propagator
    leads to
\bea\label{electron-field}
        A_0=-\0{e}{m_{\gamma ,P}^2}\rho _e,
\eea
from which the effective chemical potential $\mu _e^*$ is
    defined with a gauge invariant manner\refr{Maruyama2004}
\bea\label{chemical}
    \mu_e^*&&\equiv\mu_e+\mu _{e,I}
    =\mu _e-\0{e^2}{m_{\gamma, P}^2 } \rho _e.
\eea The $n_e(\mu ^*_e, T )$ and $\overline{n_e}(\mu ^*_e, T )$
are
    the distribution functions for (anti-)particles with $E_e =\sqrt{{\bf k}^2+m_e^2}$.
Throughout this paper, we set $m_e=0$ as done in the literature.

\subsection{Mass gap multiplier with Thomson stability}
The mass gap $m_{\gamma ,P}^2$ is a Lagrange multiplier that
    enforces relevant physical dynamics constraints.
The remaining task is how
    to determine this gap parameter reflecting the quantum fluctuating
    effects.

According to the general density correlation theory of statistical
physics, the following identity should be fulfilled at the
infrared singular critical point with strong fluctuation\bea
\0{\pa P_e}{\pa \rho _e}|_T=0.\eea Therefore the total lepton
number density has to make the system stable by minimizing the
thermodynamical potential. We find that this crucial relation can
be used to determine the unknown mass gap multiplier. Minimizing
the thermodynamical potential of the leptons system(made of
electrons and implicit
    neutrinos) stabilized by the external baryon background with the mathematically well defined effective potential
    or pressure
\eq{potential} functional with
\eq{electron-field}\refr{gulminelli2005} \bea\label{stability}
     \0{\pa\Omega _e}{\pa \rho_e}|_{m_{\gamma ,P}^2,T}=0,
\eea one has \bea\label{debye-mass-relation}
    m_{\gamma,P}^2=-\0{e^2}{3}(T^2+\0{3 {\mu ^*_e }^2}{\pi ^2})
=-m_{\gamma,e}^2, \eea which is the negative gauge invariant
electric screening mass squared obtained through the photon
polarization
    \eq{self-energy} with the full electron propagator \eq{chemical}.

With some algebra operations, it is easy to find that
\eq{stability} is also equivalent to the additional global $U(1)$
lepton conservation \bea\label{lepton-conservation}
     \0{\pa\Omega _e}{\pa \mu_e}|_{m_{\gamma ,P}^2,T}=0.
\eea The physical meanings of \eq{thermal} and
\eq{lepton-conservation} are quite different from each other. The
former is thermodynamics relation identity, while the latter is
the ``mechanical" stability criterion. The relation between
$m_{\gamma,P}^2(\mu ^*_P)$ and
        $m_{\gamma,e}^2(\mu _e^*)$ can be confirmed by comparing
        \eq{electron-field} with that obtained for protons $
        A_0={e}/{m_{\gamma ,e}^2}\rho _P $, by noting $\rho _e=\rho _P$ for the classical thermodynamics ground
    state.

The \eq{debye-mass-relation} is
    similar to the relation between the Debye mass and pressure for
    the ideal degenerate/quasi-particle gas derived with Ward-Identity
    responsible for the current conservation/gauge invariance\refr{kapusta1989}.
The magnitude of the gauge invariant
    ratio $m_{\gamma,(e,P)}^2/e^2$ is the density of states for symmetric two component degrees of freedom with spin down and up.
The contribution $\mu _{e,I}$to the effective
    chemical potential characterizes the long range correlation physics, which leads to the phase space deformation of the
    particle distribution functions.

The thermodynamics properties can be learned from the
    underlying effective potential consequently. With \eq{potential}
    and the standard thermodynamics relation
\bea \epsilon _e=\01V \0{\pa (\beta \Omega_e)}{\pa \beta }+\mu _e
\rho _e, \eea
    the energy density functional is
\bea
    \epsilon _e&&=\0{e^2}{2m_{\gamma ,P}^2} \rho _e^2 +2 \int_k E_e \[n_e (\mu _e^*,T)+{\bar n_e}(\mu
    _e^*,T)\].
\eea
\subsection{Negative pressure contribution} We analyze concretely the pressure $P_e=-\Omega_e/V$ at
$T=0$, \bea\label{pressure1}
    P_e=\0{e^2\mu_e ^6}{18 \pi ^4m_{\gamma, P}^2}+\0{\mu ^4_e}{12
    \pi^2}.
\eea With \eq{debye-mass-relation}, the pressure \eq{pressure1}
can be reduced to \bea\label{pressure}
   P_e=\013 \0{\mu ^4_e}{12 \pi^2}=\0{\mu ^4_e}{12 \pi^2}\(1-286.9\0{\alpha }{\pi }\),
\eea in order to be compared with the analytical result expected
by Kapusta in addressing the relevant compact object
    thermodynamics such as for the white dwarf stars mass correction
    resulting from the novel EM interactions of collective electrons.
This non-perturbative calculation can readily approach to his
    anticipation about a very large negative contribution to pressure
    (the coefficient of the $\alpha/\pi $ term in \eq{purbative} could
    have been $\sim 10^2$ {\em instead of} $\032$\refr{kapusta1989}),
    but has no effect on the beautiful Chandrasekhar mass limit
    because this contribution has been exactly couneracted by the opposite charged system with strongly repulsive long range correlations.

What new physics can be found? Different from the weak coupling
perturbative resummation result\refr{kapusta1989}
\bea\label{purbative}
    \0{\mu ^4_e}{12
    \pi^2}\(1-\032\0{\alpha }{\pi }\),
\eea
the analytical result \eq{pressure} does not reply on the
    coupling constant $\alpha={e^2}/({4\pi})$.
The long range quantum
    fluctuation/correlation effects leads to the negative
    contribution, which is of universal characteristic. The induced
    negative interaction energy between the collective electrons can
    be as large as their kinetic energy and the transport coefficients
    such as the shear viscosity of electrons will be reduced
    significantly compared with the ideal/quasi-particle gas model
    although it is very hard to say electrons are in a
    superconductivity state.

\section{Strongly coupling limit physics and discussions}\label{sec4}
It is very interesting that the instantaneous strong Coulomb
interaction between the like-charged
    electrons can be attractive in the dilute strongly coupling limit with the nuclear confinement effects.
This can be confirmed from the negative scattering lengths
    between electrons obtained from the oscillatory stochastic Coulomb
    potential ${e^2}{\cos(|m_{\gamma, P}|r)}/{(4\pi r)} $ with the low
    energy Born approximation.
The static Coulomb force between the itinerant electrons is
    modulated by the opposite charged nonlinear space-like baryon
    system, and so the electrons become highly turbulent and are
    confined to some extent. Like the effective fermion mass concept,
    the contribution $\mu _{e,I} $ to physical effective chemical
    potential characterizes the collective effects.
This effect is
    more obvious by rewriting $\mu ^*_e=\sqrt{m_e{*^2}+\mu _e^2 }$ at
    $T\rightarrow 0$.
Assuming the global chemical potential $\mu_e$ fixed, the
    electrons will be effectively more heavy analogous to the Kondo
    physics\refr{Coleman2005}.

Especially, the physics explored is quite similar to the quantum
    Ising characteristic BEC-BCS crossover thermodynamics studied
    with the Feshbach resonance in the opposite non-relativistic limit.
Motivated by the underlying homology\refr{chen2006}, the
analytical universal factor is calculated to be $
\xi={(d-2)(d+1)}/{d^2}$ for $d$ spatial dimensions through this
counterterm analytical method. The result for $d=3$ is exactly
consistent with the Monte Carlo study\refr{physics/0303094}. Here,
the $\xi$ is the ratio of binding energy density in the strongly
coupling unitarity limit to that for the
    non-interacting ideal fermion gas.

The surprising macroscopical like-charged attraction phenomena can
    be attributed to the hydrodynamic background effect\refr{Squires2000}
    and the similar possibility is an interesting topic in current
    cosmology etc.\refr{jenkins2004}.
Here, this counterintuitive quantum phenomena through an
    oscillating potential between the collective electrons is found from the point
    of view of the photon isospin mixing effects.

Before closing the discussion, let us stress that this is the
quantum many-body effect, i.e, it is the density
correlation/fluctuation leads to a refreshing long distance
quantum
    dragging picture. Due to the presence
of finite chemical potential in addition to the temperature
\eq{self-energy}, the Lorentz symmetry gets spontaneously broken
in the medium which leads to the nontrivial mixing of vector and
scalar although the rotation is still a good symmetry by noting
that the spin angular momentum is a good quantum number. The
nontrivial quantum transport of the strongly coupled electrons
resulting from the derived effective attractive optical potential
in the specific dilute pasta phase deserves to be further studied
in detail.

\section{Summary}\label{sec5}

Mathematically, this two-photon counterterm approach is analogous
    to the general stability principle of the Newton's action-reaction
    law in mechanics.
The screening corresponds to action, while antiscreening to
    reaction.
Rewriting $0$ as $0=\012 m_\gamma ^2 A_\mu A^\mu-\012 m_\gamma
        ^2 A_\mu A^\mu=\012 m_{\gamma,e}(\mu^*_e,T_e) ^2 A_\mu A^\mu+\012
    m_{\gamma,P}(\mu^*_P,T_P) ^2 A_\mu A^\mu$
    and taking a bulk system as two opposite charged interacting
    subsystems through an
    external electrostatic potential contribute to
    gauging/removing the infrared singularity divergences.
The external source
    formalism makes it possible to deal with the interior long range
    fluctuation effects,
    while the perturbative Furry's
    theorem limit is evaded. In spirit,
    this approach is similar to the multi-grand canonical ensemble
    statical physics, which guarantees the physical constraint conditions,
    i.e., the gauge invariance, unitarity
    and the thermodynamics self-consistency.

In conclusion, we have analyzed the isospin mixing due to photon
    in the relativistic nuclear theory and take a two-photon method
    with HLS formalism to approach the strongly correlated charged
    multi-components fermion thermodynamics by taking the mysterious
    protons, electrons and photons as an example.
The instantaneous RHA approximation and RPA techniques capture
    many of the non-trivial emergent phenomena which are crucial for
    understanding the collective effects in the strongly interacting
    charged systems.
From the pure academic viewpoint of in-medium isospin and
    spontaneous Lorentz violation effects, the EM photon's role deserves to be further explored
    because ``More is different" as said by P.W.
    Anderson\refr{Coleman2005}.
This provides a practicable counterterm compensatory scheme to
    detect and gauge the infrared EM instability in a dilute strongly
    coupling system with either vanishing or nonvanishing net global electric charge
    chemical potential.
The author acknowledges the discussions
    with Profs. Bao-an Li, Jia-rong Li, Hong-an Peng, Fan Wang and Lu Yu.
    He is also grateful to the beneficial communications
    with Profs. Ling-Fong Li and  J. Piekawewicz. Supported in part by
    the starting research fund of
    CCNU and NSFC under grant No 10675052. Initialed with the referees' suggestions for
    \refr{chen2005}.


\begin{thebibliography}{99}
\bibitem{note2}
If mistaking the nuclear system saturation condition by simply
setting $P_e=0$ of the inverse Thomson Problem at the empirical
normal density, one can have the not gauge invariant but primary
estimation $m_\gamma
    =\sqrt{2/3}{ek_f}/{\pi}=20.15 MeV$.
Instead, the medium dependent multipler parameter given by
\eq{debye-mass-relation} is the gauge invariant Debye mass of the
opposite charged lepton background. At
   $T=0$, this is Eqs.(\ref{electron-mass1}-\ref{electron-mass}) exactly confirmed with the
  self-energy calculation.
\end{thebibliography}
\end{document}